\documentclass[11pt,twoside]{article}

\usepackage{asp2006}
\usepackage{epsf}
\usepackage{lscape}

\usepackage{natbib}
\begin{document}
\title{MegaPipe: the MegaCam image stacking pipeline} 
\author{Stephen D. J. Gwyn} 
\affil{Canadian Astronomical Data Centre, Herzberg Institute of Astrophysics, 5071 West
Saanich Road, Victoria, British Columbia, Canada V9E 2E7} 

\begin{abstract} 
This paper describes the MegaPipe image processing pipeline at the
Canadian Astronomical Data Centre (CADC). The pipeline takes multiple images
from the MegaCam mosaic camera on CFHT and combines them into a single
output image. MegaPipe takes as input detrended MegaCam images and
does a careful astrometric and photometric calibration on them. The
calibrated images are then resampled and combined into image
stacks. MegaPipe is run on PI data by request, data from large surveys
(the CFHT Legacy Survey and the Next Generation Virgo Survey) and
all non-proprietary MegaCam data in the CFHT archive. The stacked
images and catalogs derived from these images are available through
the CADC website. Currently, 1500 square degrees have been
processed.

\end{abstract}

\section{Introduction}

The MegaCam data archive at the CADC is one of the largest
astronomical imaging archives in the world. While smaller in sky
coverage than the Sloan Digital Sky Survey (SDSS), the higher resolution of 
MegaCam means that
the data volume is many times greater.  The biggest barrier to using
archival MegaCam images is the effort required to process them.
MegaCam is a 36 CCD mosaic camera on CFHT \citep{megacam}.  Each
MegaCam image is about 0.7\,GB (in 16-bit integer format), making image
retrieval over the web tedious. Because of the distortion of the
MegaPrime focal plane, the images must be resampled. This involves
substantial computational demands both in terms of CPU and disk space.
While the barriers to using archival MegaCam data are not
insurmountable, they make using these data considerably less
attractive. MegaPipe aims to increase usage of MegaCam data by
removing these barriers by processing all public MegaCam data,
combining the multiple individual input images into a single, well-calibrated, 
distortion-free output image. This paper briefly describes
the MegaPipe image processing pipeline.  MegaPipe is discussed in
greater detail in \citet{gwyn2008}.

\section{Input Image Quality Control and Image Grouping}
\label{sec:group}
The first step of the MegaPipe pipeline is to ensure that each
input image can be calibrated.  Images with short exposure times
($<$50 seconds), images of nebulae or images taken under conditions of
poor transparency cannot be used.  In order for photometric
calibration to take place, an image must either be taken on a
photometric night or contain photometric standards, either from the SDSS
\citep{sdss} or sources from previously processed MegaPipe images.
Each image is inspected visually. Images with obviously asymmetric
PSFs (due to loss of tracking), unusually bad seeing,  bad focus
or other major defects are discarded.
Images taken at similar pointings on the sky are then grouped together.  In
the case of a particular project like the CFHTLS, this grouping is
obvious; related images are known {\it a priori}.  When processing the
rest of the MegaCam archival images, MegaPipe scans through the
database of MegaCam centers and groups neighboring images using a
friends-of-friends algorithm.  Two images are deemed ``friends'' if
their centers lie within 0.1\deg. It is generally not worth stacking
less than four images, so only groups with at least four images in one
filter are queued for processing.

\section{Astrometric Calibration}

The first step in the astrometric calibration pipeline is to run the
well-known SExtractor \citep{hihi} source detection software on each
image. The SExtractor catalog is cleaned of cosmic rays and extended
sources to leave only real objects with well-defined centres: stars
and (to some degree) compact galaxies.

For the first band to be reduced, these observed source catalogs are
matched with an external astrometric catalog to provide an
initial astrometric solution. This external catalog is either the
USNO A2 or the SDSS Data Release 7 (DR7).  For the subsequent
bands, the image catalogs are first matched to the USNO to provide a
rough WCS and then matched to a catalog generated using the first
image so as to precisely register the different bands.

The higher order terms of the astrometric solution are determined on
the scale of the entire mosaic; the distortion map of the entire focal
plane is measured. This distortion is well described by a polynomial
with second- and fourth-order terms in radius measured from the centre
of the mosaic. Determining the distortion in this way means that only
two parameters need to be determined (the coefficients of $r^2$ and
$r^4$) with typically (20-50 stars per chip) $\times$ (36 chips)
$\simeq$ 1000 observations. If the analysis is done chip-by-chip, a
third-order solution requires (10 parameters per chip) $\times$ (36
chips)= 360 parameters. This is less satisfactory.

>From the global distortion, the distortion local to each CCD is
determined. The local distortion is translated into a linear part
(described by the CD matrix) and a higher order part (described by the
PV keywords). The higher order part is third order as well, but the
coefficients depend directly and uniquely on the two-parameter global
radial distortion.

\section{Photometric Calibration}

The SDSS DR7 serves as the basis of the
photometric calibration.  All images lying in the SDSS can be directly
calibrated without referring to other standard stars.  The catalog for
each MegaCam image is matched to the corresponding catalog from the
SDSS.  The difference between the instrumental MegaCam magnitudes and
the SDSS magnitudes (converted to the MegaCam system) gives the
zero-point for that exposure or that CCD. The zero-point difference is
determined by median, not mean. There are about 10,000 SDSS sources per
square degree, but when one cuts by stellarity and magnitude this
number drops to around 1,000. It is best to only use the stars (the
above colour terms are more appropriate to stars than galaxies) and to
only use the objects with $17<{\rm mag}<20$ (the brighter objects are
usually saturated in the MegaCam image and including the fainter
objects only increases the noise in the median). This process can be
used for data from any night, photometric or not.

For objects outside the SDSS, the Elixir \citep{elixir} photometric keywords are
used, with modifications. The Elixir zero-points were compared to
those determined from the SDSS using the procedure above for a large
number of images. There are systematic offsets between the two sets of
zero-points, particularly for the U-band. These offsets show
variations with epoch, which are caused by modifications to the Elixir
pipeline. For MegaPipe, the offsets are applied from the Elixir
zero-points to bring them in line with the SDSS zero-points.

\section{Image Stacking and Catalog Generation}

The calibrated images are coadded using the program SWarp
\citep{swarp}.  The resulting stacks are simple FITS files (not
multi-extension FITS files) measuring about 20000 pixels by 20000
pixels (about 1 degree by 1 degree), depending on the input dither
pattern, and are about 1.7\,GB in size. They have a sky level of 0
counts. They are scaled to have a photometric zero-point of 30.000 in
AB magnitudes.  A weight map (inverse variance) of the same size is
also produced.

SExtractor is run on each stack using the weight map.  The resulting
catalogs only pertain to a single band; no multi-band catalogs
have been generated, except for special cases, like the CFHTLS. While
this fairly simple approach works well in many cases, it is probably
not optimal in some situations. Depending on the application, some
users may wish to run their own catalog generation software on the
stacks.

\section{Checks on Astrometry and Photometry}

The positions and magnitudes of the objects in the catalogs are used
to check the astrometric and photometric uncertainties.  Catalogs from
different bands of the same group are matched up. The residuals
indicate the internal astrometric uncertainty is typically
0.025\arcsec. Similarly, the astrometric residuals between
independently produced overlapping catalogs indicate the the
repeatability is 0.06\arcsec.  Examining the residuals between the
catalogs and the astrometric reference catalogs (either the USNO or
the SDSS) give an external astrometric accuracy of 0.2\arcsec, after
allowing for the uncertainties in the astrometric reference catalogs
themselves.

The photometric accuracy is assessed by comparing catalogs from
different pointings and measuring the systematic offset. The average
offset was found to be 0.015 magnitudes. This only measures the
repeatability; the ultimate accuracy is limited by the photometric accuracy
of the SDSS and the uncertainty in converting from SDSS to MegaCam
magnitudes or the accuracy of the Elixir calibration. Analyses
of both methods yield accuracies of 0.02-0.03 magnitudes.

The limiting magnitude of the images is assessed in three ways:
locating the peak of the number counts, locating the $5\sigma$
detection limit by finding the faintest object whose magnitude error is
0.198 mag or less, and finally by adding fake objects to a subsection
of the image. This last method also assesses the surface brightness
limit as well as the magnitude limit.

\section{Production and Distribution}

The CFHT data access policy is such that large chunks of data become
public on a semi-annual basis. Therefore, every six months, the grouping
algorithm described in Section \ref{sec:group} is run to select new
groups of images for stacking.  (PI programs are processed on a
case-by-case basis.) MegaPipe is run on the CADC processing grid,
using Sun Grid Engine as a scheduler. Input images are retrieved from
the MegaCam archive, processed and the output MegaPipe images are
inserted back into the archive. Manual intervention occurs at the
beginning (visually inspecting the input images) and the end
(examining various diagnostics of the astrometric and photometric
calibrations and spot checking of the images themselves).  The
processing itself is completely autonomous. The CADC processing
cluster can run MegaPipe on six months worth of MegaCam images
(typically 300 groups) in three days.

The MegaPipe images and catalogs are distributed via the CADC
website.\footnote{\tt http://www.cadc.hia.nrc.gc.ca/megapipe/index.html} The archive is
searchable via a conventional query page as well as a graphical
interface (based on the Google Maps API) described elsewhere in these
proceedings. In addition an image cutout service is provided.

\acknowledgements This research used the facilities of the Canadian Astronomy Data
Centre operated by the National Research Council of Canada with the
support of the Canadian Space Agency.

Based on observations obtained with MegaPrime/MegaCam, a joint project
of CFHT and CEA/DAPNIA, at the Canada-France-Hawaii Telescope (CFHT)
which is operated by the National Research Council (NRC) of Canada,
the Institute National des Sciences de l'Univers of the Centre
National de la Recherche Scientifique of France, and the University of
Hawaii.

\end{document}